\documentclass[aps,prb,twocolumn,groupedaddress,showpacs]{revtex4}
\usepackage{graphicx}

\begin{document}
\title{Ferromagnetism and giant magnetoresistance in
the rare earth fullerides Eu$_{6-x}$Sr$_x$C$_{60}$}
\author{Kenji Ishii}
\email{kenji@spring8.or.jp}
\affiliation{Department of Physics, School of Science,
The University of Tokyo, 7-3-1 Hongo, Bunkyo-ku, Tokyo 113-0033, Japan}
\affiliation{Synchrotron Radiation Research
Center, Kansai Research Establishment, Japan Atomic Energy Research
Institute, 1-1-1 Kouto Mikazuki-cho Sayo-gun, Hyogo 679-5148, Japan}
\author{Akihiko Fujiwara}
\altaffiliation[Present address: ]{School of Materials Science,
Japan Advanced Institute of Science and Technology, 1-1 Asahidai,
Tatsunokuchi, Ishikawa 923-1292, Japan}
\affiliation{Department of Physics, School of Science,
The University of Tokyo, 7-3-1 Hongo, Bunkyo-ku, Tokyo 113-0033, Japan}
\author{Hiroyoshi Suematsu}
\altaffiliation[Present address: ]{Material Science Division, SR
Research Laboratory, Japan Synchrotron Radiation Research Institute,
1-1-1 Kouto Mikazuki-cho, Sayo-gun, Hyogo 679-5198, Japan}
\affiliation{Department of Physics, School of Science,
The University of Tokyo, 7-3-1 Hongo, Bunkyo-ku, Tokyo 113-0033, Japan}
\author{Yoshihiro Kubozono}
\affiliation{Department of Chemistry, Faculty of Science,
Okayama University, 3-1-1 Tsushima-naka, Okayama 700-8530, Japan}
\date{\today}

\begin{abstract}
We have studied crystal structure, magnetism and electric transport
properties of a europium fulleride Eu$_6$C$_{60}$ and its
Sr-substituted compounds, Eu$_{6-x}$Sr$_x$C$_{60}$. They have a $bcc$
structure, which is an isostructure of other $M_6$C$_{60}$ ($M$
represents an alkali atom or an alkaline earth atom). Magnetic
measurements revealed that magnetic moment is ascribed to the divalent
europium atom with $S$ = 7/2 spin, and a ferromagnetic transition was
observed at $T_C$ = 10 - 14 K. In Eu$_6$C$_{60}$, we also confirm the
ferromagnetic transition by heat capacity measurement. The striking
feature in Eu$_{6-x}$Sr$_x$C$_{60}$ is very large negative
magnetoresistance at low temperature; the resistivity ratio $\rho$($H$
= 9 T)/$\rho$($H$ = 0 T) reaches almost 10$^{-3}$ at 1 K in
Eu$_6$C$_{60}$. Such large magnetoresistance is the manifestation of a
strong $\pi$-$f$ interaction between conduction carriers on C$_{60}$
and 4$f$ electrons of Eu.
\end{abstract}


\maketitle

\section{INTRODUCTION}
Since the discovery of fullerenes, C$_{60}$ compounds have given us
various opportunities for the research in condensed matter physics and
materials science. Much attention was attracted to the
superconductivity in $A_3$C$_{60}$ ($A$ is an alkali atom)
\cite{Hebard1}. As for the magnetism, TDAE-C$_{60}$ (TDAE is
tetrakisdimethylaminoethylene) shows a ferromagnetic transition
\cite{Allemand1}, while antiferromagnetic (or spin density wave) ground
state was observed in polymeric $A_1$C$_{60}$ \cite{Chauvet1},
Na$_2$Rb$_{0.3}$Cs$_{0.7}$C$_{60}$ \cite{Arcon1}, and
three-dimensional (NH$_3$)$A_3$C$_{60}$ \cite{Takenobu1}. In these
compounds a magnetic moment is considered to be carried by an electron
on C$_{60}$ molecule. Because various atoms and molecules can be
intercalated into C$_{60}$ crystal, we also expect the magnetic
C$_{60}$ compounds in which magnetic moment is carried by
intercalants. In this viewpoint, rare earth metal is a good
candidate. The research of rare earth fullerides was reported for Yb
\cite{Oezdas1} and Sm \cite{Chen1} in relation to the
superconductivity, but little effort has been made to study the
magnetic properties. The only case of magnetic study in rare earth
fullerides is for europium. Europium has a magnetic moment of 7$\mu_B$
($S$ = 7/2, $L$ = 0, and $J$ = 7/2) in the divalent state, while it is
non-magnetic ($S$ = 3, $L$ = 3, and $J$ = 0) in the trivalent state. A
photoemission study of C$_{60}$ overlayered on Eu metal revealed the
charge transfer from Eu to C$_{60}$ and the formation of fulleride
\cite{Yoshikawa1}.  Ksari-Habiles {\it et al}. \cite{Ksari1,Claves1}
investigated the crystal structure and magnetic properties of
Eu$_{\sim3}$C$_{60}$ and Eu$_6$C$_{60}$; they observed some magnetic
anomalies in Eu$_6$C$_{60}$.

In this paper, we report the ferromagnetic transition of
Eu$_6$C$_{60}$, which was observed at $T_C \sim$ 12 K in magnetic and
heat capacity measurements. We also investigated the substitution
effect from Eu to non-magnetic Sr, and the ferromagnetic transition
temperature was found to change little with the Sr concentration. In the
resistivity measurement we found a huge negative magnetoresistance
below around $T_C$; the reduction ratio of resistivity
$\rho(H)/\rho(0)$ is almost 10$^{-3}$ at 1 K in Eu$_6$C$_{60}$. This
ratio is comparable to those in perovskite manganese oxides, which is
known as colossal magnetoresistance (CMR). However Eu$_6$C$_{60}$
should be categorized as a new class of giant magnetoresistive
compounds in the sense that (1) the magnitude of magnetoresistance
increases very steeply with decreasing temperature rather than the
vicinity of $T_C$, (2) the compound consists of a molecule with novel
structure. These features can open the further possibility to find a
new magnetic and magnetoresistive material.

\section{EXPERIMENTAL PROCEDURES}
Polycrystalline samples of Eu$_{6-x}$Sr$_x$C$_{60}$ were synthesized
by solid-state reaction. A stoichiometric amount of mixture of Eu, Sr
and C$_{60}$ powders, which was pressed into a pellet and sealed in a
quartz tube in vacuum, was heat-treated at 600 $^{\circ}$C for about
10 days. In the course of the heat treatment the sample was ground for
ensuring the complete reaction. Because the sample is very unstable in
air, we treated it in a glove box with inert atmosphere.

Powder x-ray diffraction experiments were carried out by using
synchrotron radiation x-rays at BL-1B in Photon Factory, KEK,
Tsukuba. The samples was put into a glass capillary in 0.3 mm diameter
and an imaging plate was used for the detection \cite{Fujiwara1}.
Magnetic measurements were performed using a SQUID magnetometer. In
the heat capacity measurement by relaxation method, the sample was
pressed into a pellet and sealed by grease to keep from exposure to
air. Eu $L_{III}$-edge XANES (x-ray absorption near edge structure)
was measured in the fluorescence method at BL01B1 of SPring-8,
Harima. The resistivity measurements were carried out by the 4-probe
method.  Four gold wires were attached to a pressed pellet of
polycrystalline sample with sliver paste. The sample was put into a
capsule and sealed in He atmosphere.

\section{RESULTS}
\begin{figure*}
\includegraphics{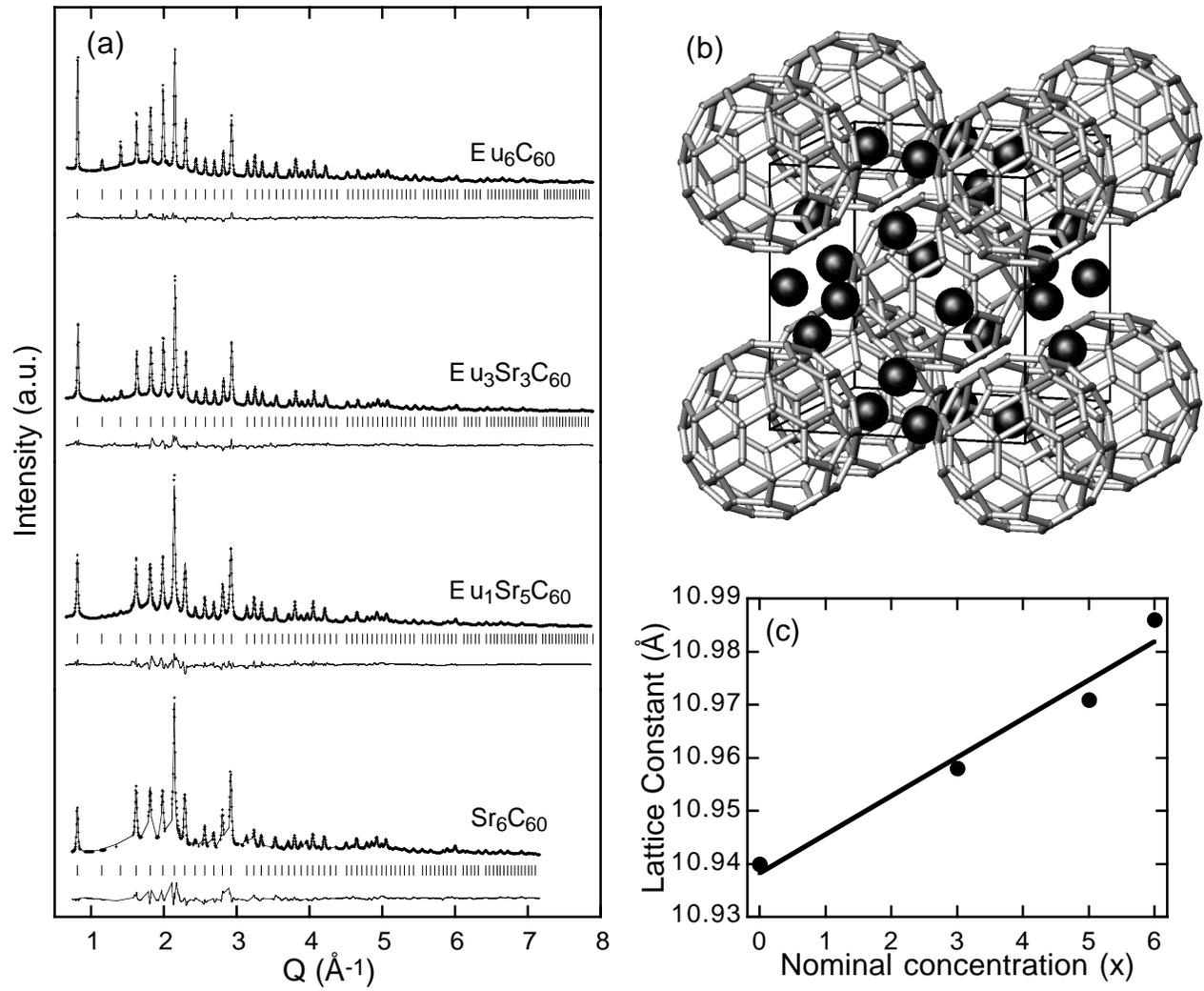}
\caption{\label{fig:structure} (a) X-ray diffraction spectra of
Eu$_{6-x}$Sr$_x$C$_{60}$. The wavelength of x-ray is 0.8057 \AA\ for
$x$ = 0, 3, 5, and 0.8011 \AA\ for $x$ = 6. (b) Schematic view of the
crystal structure of Eu$_{6-x}$Sr$_x$C$_{60}$. The black ball
represents a metal atom. (c) Lattice constant vs. nominal
concentration of Eu ($x$). The lattice constant changes linearly with
$x$, following the Vegard's law.}
\end{figure*}

X-ray diffraction spectra of Eu$_{6-x}$Sr$_x$C$_{60}$ are shown in
Fig.\ \ref{fig:structure}(a). The spectrum of Sr$_6$C$_{60}$ is also
presented as a reference. The wavelength of x-ray is 0.8057 \AA\ for
$x$ = 0, 3, 5, and 0.8011 \AA\ for $x$ = 6.  They all can be
understood by a $bcc$ structure which is an isostructure of other
$M_6$C$_{60}$ in alkali \cite{Zhou1}, alkaline earth\cite{Kortan1} and
rare earth (Sm) \cite{Chen2} fullerides. The Rietveld refinements
based on the space group $Im\overline{\it 3}$ were performed with use
of the RIETAN program \cite{Izumi1,Kim1}.  In the refinements, only
two atomic coordinates ($x$ for C1 and C3) are refined in C$_{60}$
molecule, which corresponds to the refinement of the length of 6:6
bond (the bond between two hexagons) and 5:6 bond (the bond between
hexagon and pentagon). In the compounds of $x$ = 3 and 5, the sum of
the metal concentration is fixed to unity. The results of refinement
are presented in Table \ref{tab:structure} and obtained structure is
shown in Fig.\ \ref{fig:structure}(b). This crystal structure of
Eu$_6$C$_{60}$ is consistent with the previous works
\cite{Claves1,Ootoshi1}, but we observed little trace of the secondary
phase in the present sample. In the Sr-substituted compounds, the
values of Eu concentration are in good agreement with the nominal
ones.

As seen in Fig.\ \ref{fig:structure}(c), the obtained lattice
constants change linearly with the nominal Eu concentration, which
means they follow the Vegard's law, and confirms the formation of
solid solution at $x$ = 3 and 5. This result is attributed to the fact
that ionic radius of Eu$^{2+}$ and Sr$^{2+}$ is quite similar, while
the substitution of Ba for Eu results in the phase separation.

\begin{figure*}
\includegraphics{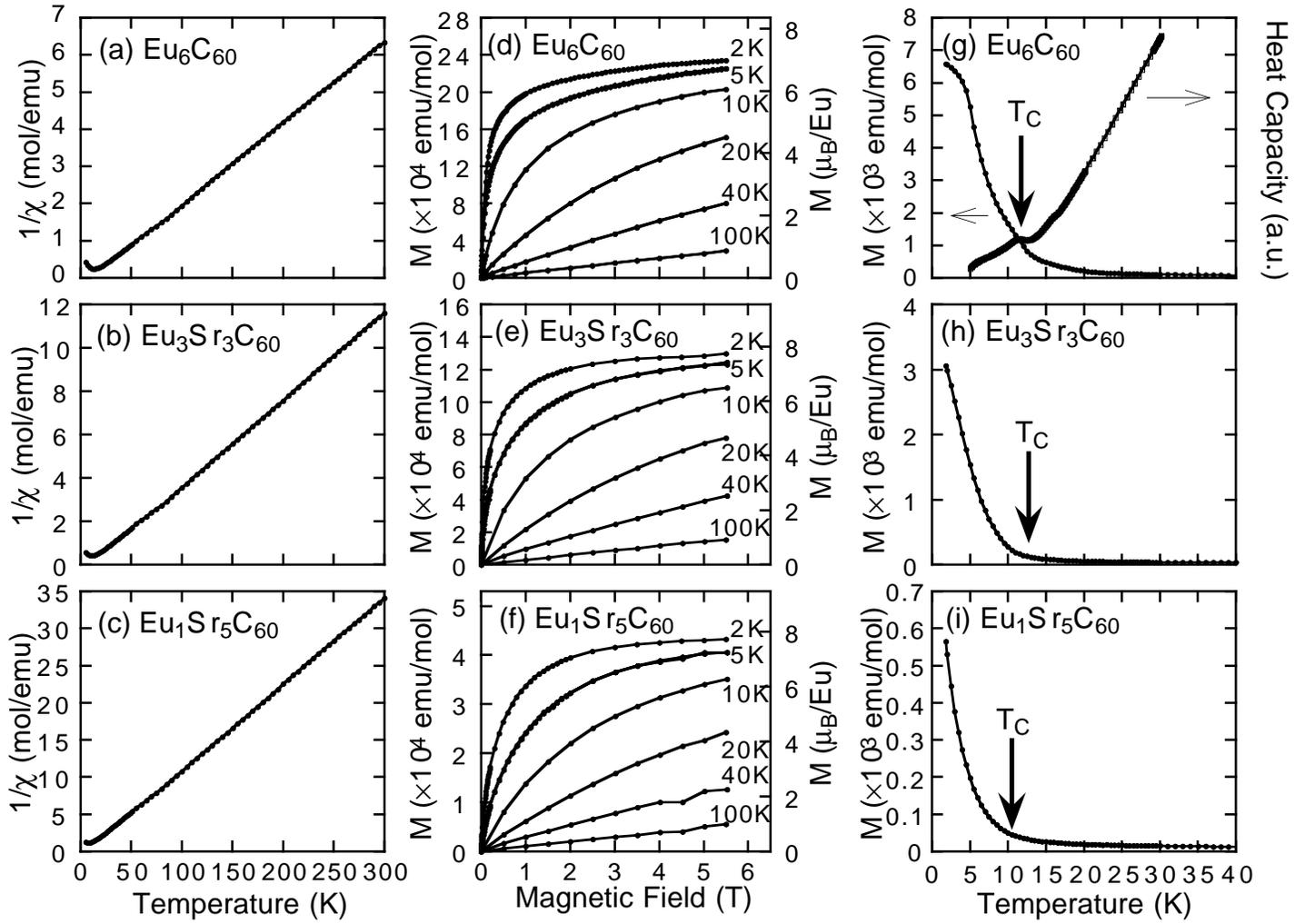}
\caption{\label{fig:magnetism} (a)-(c) Inverse magnetic susceptibility
obtained from the magnetization at 1 and 2 T. (d)-(f) Magnetization
curve. (g)-(i) Temperature dependence of the magnetization at 3
mT. Heat Capacity of Eu$_6$C$_{60}$ is also presented in (g). The
arrows indicate the ferromagnetic transition temperatures determined
the peak position of the heat capacity measurement for Eu$_6$C$_{60}$,
and estimated from the Arrott plot for Eu$_3$Sr$_3$C$_{60}$ and
Eu$_1$Sr$_5$C$_{60}$.}
\end{figure*}

\begin{figure*}
\includegraphics{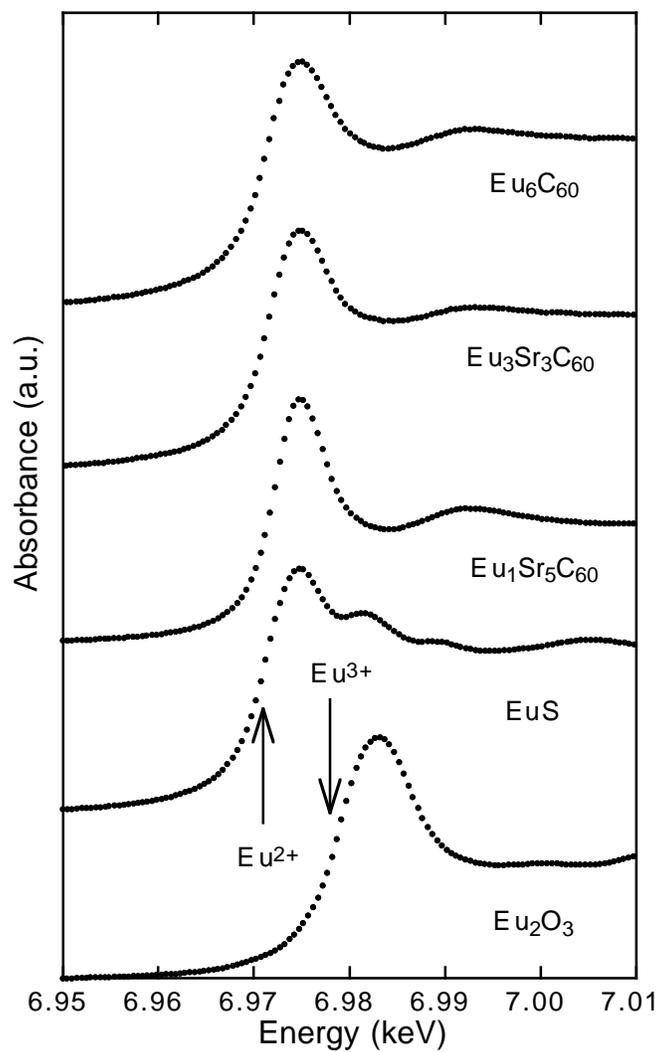}
\caption{\label{fig:xanes} XANES spectra of
Eu$_{6-x}$Sr$_x$C$_{60}$. The arrows indicate the absorption edge in
divalent and trivalent reference.}
\end{figure*}

Figures \ref{fig:magnetism} show the result of magnetic measurements
of Eu$_{6-x}$Sr$_x$C$_{60}$. Above 30 K, magnetic susceptibility
($\chi$) follows the Curie-Weiss law, as shown in Figs.\
\ref{fig:magnetism}(a)-(c). The effective Bohr magneton estimated from
Curie constant and the Weiss temperature are summarized in Table
\ref{tab:magnetism} \cite{eumag}. The former agrees with the Eu$^{2+}$
state ($S$ = 7/2, $L$ = 0, and $J$ = 0). The field dependence of
magnetization at 2 K gives the saturation moment close to 7$\mu_{B}$,
which is consistent with the magnetic moment of Eu$^{2+}$. Moreover
the Eu$^{2+}$ state has been also confirmed by Eu $L_{III}$-edge XANES
experiments, as seen in Fig.\ \ref{fig:xanes}. The spectra of EuS and
Eu$_2$O$_3$ was also presented as a reference of divalent and
trivalent of Eu, and absorption edges of Eu$_{6-x}$Sr$_x$C$_{60}$ are
very close to that of EuS. The divalent state of Eu also observed in
the case that Eu atom exists inside the C$_{60}$ cage, namely,
metallofullerene Eu@C$_{60}$ \cite{Inoue1}.

Temperature dependence of magnetization at a weak field of 3 mT
(Figs.\ \ref{fig:magnetism}(g)-(i)) shows a steep increase of
magnetization below 10-14 K, indicating a ferromagnetic transition. To
confirm the presence of the ferromagnetic phase transition, we
measured heat capacity for Eu$_6$C$_{60}$. In Fig.\
\ref{fig:magnetism}(g) we show the temperature dependence of heat
capacity including that of grease. An obvious peak can be seen near
the transition temperature, which is an evidence of the ferromagnetic
phase transition.  The $T_C$ is determined to be 11.6 K from the peak
position. We ascertained that there was no anomaly in specific heat in
this temperature region for grease, which was used to keep the sample
from exposure to air. We can also see a smaller peak near 16 K, whose
origin has not been clarified yet, but we consider that it does not
come from a magnetic origin because of no anomaly in the temperature
dependence of magnetization. The transition temperatures for
Eu$_3$Sr$_3$C$_{60}$ and Eu$_1$Sr$_5$C$_{60}$ are estimated from the
Arrott plot \cite{Arrott1} at 12.8 K and 10.4 K, respectively. These
values are very close to the Weiss temperature mentioned above. In
Eu$_6$C$_{60}$, the transition temperature estimated from the Arrott
plot is a little larger value (13.7 K) than that from the heat
capacity measurement, but it is not so important in the following
discussion.

From these evidences we conclude that Eu$_{6-x}$Sr$_x$C$_{60}$ shows a
ferromagnetic transition at $T_C$ = 10-14 K, and the magnetic moment
is ascribed to Eu$^{2+}$. In the previous work of Eu$_6$C$_{60}$,
Ksari-Habiles {\it et al}. \cite{Ksari1} observed a mixed valence
state of Eu (Eu$^{2+}$ and Eu$^{3+}$) and three successive magnetic
anomalies, which is different from the present work; a possible reason
is that their sample might contain a secondary phase other than
Eu$_6$C$_{60}$.

\begin{figure*}
\includegraphics{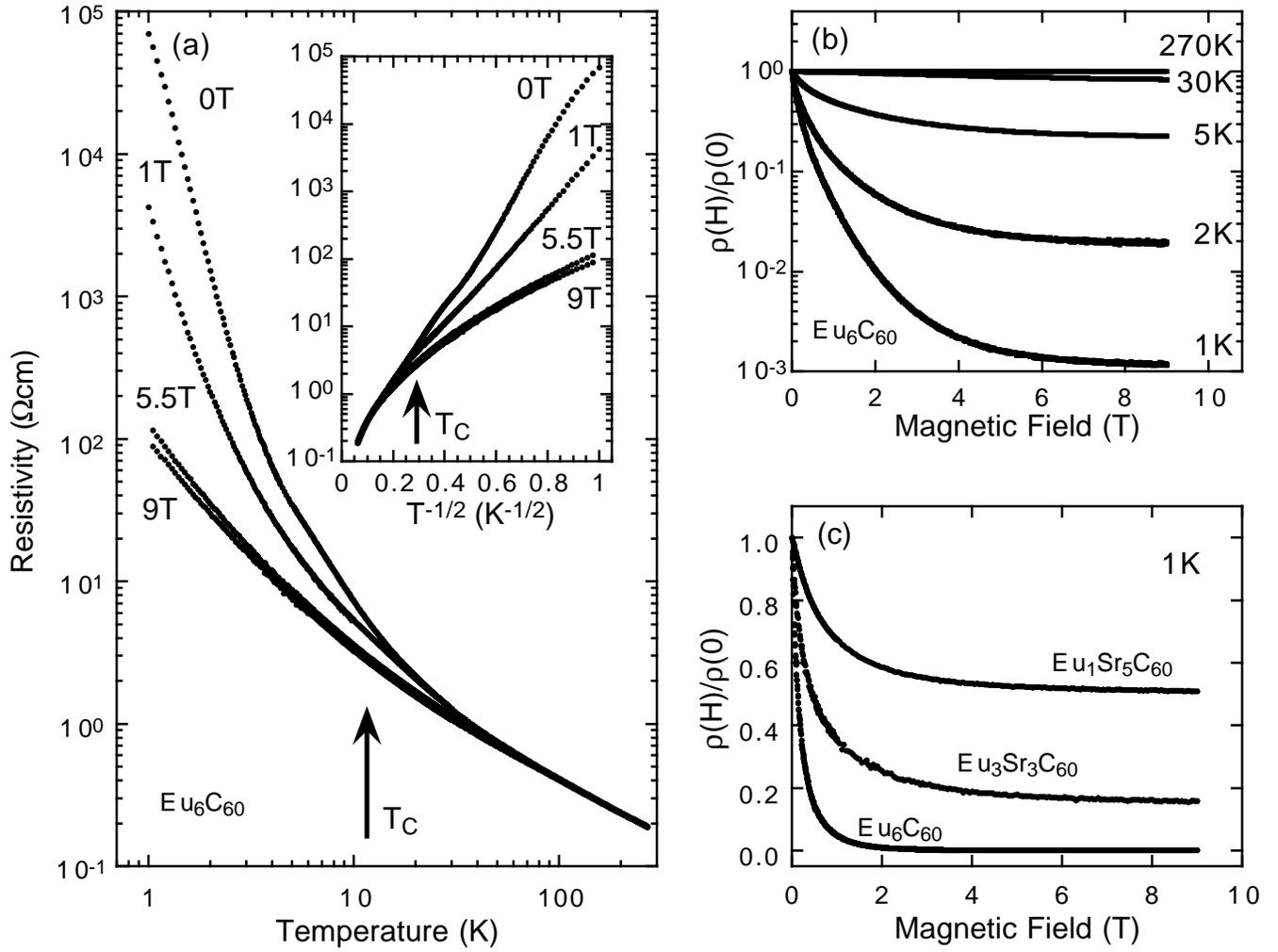}
\caption{\label{fig:resistivity} (a) Temperature dependence of
resistivity of polycrystalline Eu$_6$C$_{60}$ at some magnetic
fields. The arrow indicates the ferromagnetic transition
temperature. (b) Magnetic field dependence of resistivity of
Eu$_6$C$_{60}$ normalized at zero field for some temperatures. (c)
Magnetic field dependence of resistivity of Eu$_{6-x}$Sr$_x$C$_{60}$
normalized at zero field measured at 1 K.}
\end{figure*}

Figure \ref{fig:resistivity} (a) show the temperature dependence of
electric resistivity of Eu$_6$C$_{60}$ measured at some magnetic
fields. A most striking feature in resistivity is the huge negative
magnetoresistance below around $T_C$. The negative magnetoresistance
becomes much more significant at lower temperature. In the case of
Eu$_6$C$_{60}$, magnetoresistivity $\rho$($H$ = 9 T) is three orders
magnitude smaller than $\rho$($H$ = 0 T) at 1 K, as seen in Fig.\
\ref{fig:resistivity} (b). This large negative magnetoresistance is
comparable to those of the colossal magnetoresistance (CMR) materials
such as perovskite manganese oxides, where CMR effect is seen only
near the ferromagnetic transition temperature. We also observed a
relatively large magnetoresistance in the Sr-substituted compounds, as
shown in \ref{fig:resistivity} (c). There is no difference between
magnetoresistances in the transverse ($H \perp I$) and longitudinal
($H \parallel I$) configurations ($I$ represents electric current),
suggesting the magnetoresistance in Eu$_6$C$_{60}$ is not ascribed to
the orbital motion of free carriers.

\section{DISCUSSION}
Such giant magnetoresistance is a manifestation of the strong
interaction between conduction carriers and localized magnetic
moments; namely, the strong $\pi$-$f$ interaction exists in
Eu$_6$C$_{60}$. When we consider formal valence state of
(Eu$^{2+}$)$_6$C$_{60}^{12-}$, $t_{1g}$ band of C$_{60}$ is completely
filled and the compound should become an insulator. In this case, the
interaction between conduction carrier and localized moment is
considered to be week, assuming that the conduction carrier mainly
passes on C$_{60}$ molecules. If Eu orbitals hybridize with C$_{60}$
orbitals and form a part of conduction band, much enhancement of the
interaction must occur. In the band calculation for Sr$_6$C$_{60}$ and
Ba$_6$C$_{60}$ \cite{Saito1} which have the same $bcc$ crystal
structure and the same valence state as Eu$_6$C$_{60}$, the
hybridization of the $d$ orbital of metal atom and the $t_{1g}$
orbital of C$_{60}$ exists and make the compounds metallic. This fact
is confirmed experimentally \cite{Gogia1}. The hybridization is more
significant in Sr$_6$C$_{60}$ than Ba$_6$C$_{60}$ due to the smaller
lattice constant of Sr$_6$C$_{60}$. The band structure of
Eu$_6$C$_{60}$ has not been studied yet, but such hybridization of the
5$d$ and/or 6$s$ orbitals of Eu and the $t_{1g}$ orbital of C$_{60}$
is plausible in Eu$_6$C$_{60}$ because Eu$_6$C$_{60}$ has a further
smaller lattice constant than Sr$_6$C$_{60}$.

The $\pi$-$f$ interaction is likely to affect to magnetic interaction
of 4$f$ electrons and the origin of ferromagnetism in Eu$_6$C$_{60}$
may be ascribed to the indirect exchange interaction. In the $bcc$
structure, an Eu atom has 4 nearest neighbor Eu atoms (the distance
between two Eu atoms is 3.89 \AA). Therefore, in the case of
Eu$_1$Sr$_5$C$_{60}$, 5 of 6 Eu atoms are replaced by non-magnetic Sr
atoms, Eu atoms can no longer have the three-dimensional Eu network,
so that the direct interaction fails completely.  Nevertheless $T_C$
does not show a drastic change. This is a quite contrast with the case
of magnetic semiconductor EuO, where the direct exchange interaction
between Eu atoms is important \cite{Kasuya2} and the substitution of
Ca for Eu significantly reduces the ferromagnetic transition
temperature \cite{Samokhvalov1}. This fact indicates that the
ferromagnetism in Eu$_6$C$_{60}$ comes from the indirect exchange
interaction via C$_{60}$ molecules, and the $\pi$-$f$ interaction has
an important role in the present system.

Now we discuss the origin of the giant magnetoresistance. The features
of magnetoresistance in Eu$_6$C$_{60}$ are (1) negative
magnetoresistance occurs below around $T_C$, (2) saturation field of
magnetizaion is close to that of magnetoresistance ratio
($\rho(H)/\rho(0)$), as seen in the top curve of Fig.\
\ref{fig:magnetism}(d) and the bottom curve of Fig.\
\ref{fig:resistivity}(c), (3) magnetoresistance is much enhanced at
lower temperatures; The magnetoresistance ratio ($\rho(H)/\rho(0)$)
does not seem to saturate with decreasing temperature, while the
magnetization almost saturates at 2 K and 5.5 T. The feature (1)
indicates that the present MR is closely related to the ferromagnetic
transition. In usual ferromagnetic metal, spin fluctuation scatters
conduction electrons and causes negative magnetoresistance
\cite{Gennes1,Fisher1}.  This effect may be an origin of the
magnetoresistance near $T_C$, but this is not the case for the giant
magnetoresistance in Eu$_6$C$_{60}$ at lower temperature, because such
effect is remarkable in the vicinity of $T_C$ inconsistent with the
feature (3). The feature (2) suggests that magnetoresistance is
related to the magnetization.  Furthermore, when we see
$\rho(H)/\rho(0)$ in log scale, the difference of $\rho(H)/\rho(0)$
between 2 K and 1 K is almost one order (Fig.\
\ref{fig:resistivity}(b), while that of magnetization must be
small. This means there is another factor, in addtion to the
magnetization, to determine the magnetoresistance.  It is probably
temperature, that is, an activation process needs to be considered in
the origin of the magnetoresistance.

One possible interpretation of magnetoresistance in Eu$_6$C$_{60}$ is
the spin-dependent tunneling at the grain boundary \cite{Helman1}. In
the case of ferromagnetic granular metal, the conductivity is
dependent on the tunneling probability of carriers through insulating
barrier between grains, and the probability crucially depends on the
spin polarization of carriers. In this case, each grain is assumed to
be conductive and surrounded by less conductive surface. Because we
measured the resistivity in a pellet of polycrystalline sample and
Eu$_6$C$_{60}$ is very unstable in air, the surface may react to be
insulative barrier, even if the sample is treated in high purity inert
atmosphere. However we should note that the insulating region is
considered to be limited only on the thin surface because unidentified
peaks in XRD spectrum are very weak and they are considered not to
affect to the magnetic and heat capacity measurements. As shown in the
inset of Fig.\ \ref{fig:resistivity}(a), the temperature dependence of
resistivity is represented as $\rho(T) \propto \exp(T_0/T)^{1/\alpha}
(\alpha \sim 2)$, rather than the activation type which is expected in
a usual semiconductor. The value of $T_0$ is about 180 K at 0T. This
fact suggests that the resistivity in our sample might be governed by
the tunneling at the boundaries \cite{Sheng1}. Our preliminary Hall
effect measurement gives $R_H$ = +5$\times$10$^{-2}$ cm$^3$/C at 250
K, corresponding to the hole density of 1$\times$10$^{20}$ cm$^{-3}$
(0.1 hole per C$_{60}$); this means that intrinsic Eu$_6$C$_{60}$ can
have relatively high conductivity and is possibly metallic by
hybridization of the C$_{60}$ and metal orbitals mentioned above. Note
that Hall voltage is less sensitive to the grain boundary
effect. Helman and Abeles \cite{Helman1} considered the magnetic
exchange energy $E_M$ and gave the mangetoconductivity as
\begin{equation}
\label{eqn:pnot0}
\sigma(H,T) = \sigma_0\left[\cosh(E_M/2k_BT)-P\sinh(E_M/2k_BT)\right],
\end{equation}
where $P$ is the spin polarization of carrier and
$E_M$=(1/2)$J$[1-$m^2$]. $J$ is the exchange coupling constant between
a conduction carrier and a ferromagnetic metal grain and $m$ is the
magnetization normalized by the saturation value. The equation
(\ref{eqn:pnot0}) gives a negative magnetoresistance of orders of
magnitude only when $P$ is very close to unity. If $P$ = 1, we obtain
\begin{equation}
\label{eqn:pequal0}
\rho(H)/\rho(0) = \exp(-Jm^2/4k_BT).
\end{equation}
Magnetoresistance in equation (\ref{eqn:pequal0}) becomes large with
decreasing temperature, which agree qualitatively with the feature (2)
mentioned above. The assumption of $P$ = 1 might be unrealistic in
usual ferromagnetic metals. However, if the exchange interaction
between Eu atoms is accomplished via $\pi$-bands of C$_{60}$ as
discussed earlier, we can expect a large spin polarization of
$\pi$-electrons.

We can also consider the effect of magnetic polaron. In magnetic
semiconductors such as Eu chalcogenides, a carrier makes surrounding
magnetic moments be polarized via exchange interaction and forms a
magnetic polaron \cite{Kasuya1}. At zero field, magnetic polarons have
to move with flipping some magnetic moments which are more or less
randomly oriented, and their conduction is suppressed. Application of
magnetic field aligns spin directions and carriers become mobile. As a
result, negative magnetoresistance occurs. The negative
magnetoresistance above $T_C$ can be attributed to this picture. Even
in the ferromagnetic phase, magnetic moments have to be flipped at a
magnetic domain boundary for the motion of carrier. Because remnant
magnetic moment is little, as seen in Figs.\ \ref{fig:magnetism} (d),
many magnetic domains exist in our sample of Eu$_6$C$_{60}$. The
crucial point of above two interpretations (spin dependent tunneling
and magnetic polaron) are that carriers must overcome large exchange
interaction with localized spins when they go into the region of
different orientation of magnetic moments.

\section{SUMMARY}
We have measured crystal structure, magnetic properties and
magnetoresistance in polycrystalline Eu$_6$C$_{60}$ and its
Sr-substituted compounds, Eu$_{6-x}$Sr$_x$C$_{60}$. They all have a
$bcc$ structure and the compounds of $x$ = 3 and 5 form a solid
solution concerning the occupation of metal atom. A ferromagnetic
transition is observed at $T_C \sim$ 12 K in Eu$_6$C$_{60}$ and all Eu
atoms are in divalent state with a magnetic moment of 7$\mu_B$ ($S$ =
7/2). The fact that the substitution of non-magnetic Sr for Eu affects
little to $T_C$ indicates the ferromagnetic interaction is caused
through the conduction carriers.  In the resistivity measurement, we
have found that Eu$_6$C$_{60}$ showed a huge negative
magnetoresistance and $\rho(H)/\rho(0)$ reduced almost 10$^{-3}$ at
$H$ = 9 T and $T$ = 1 K. The precise mechanism of magnetoresistance
has not clarified yet, but it manifests a strong interaction between
$\pi$-conduction electrons of C$_{60}$ and 4$f$ electrons on Eu.

\begin{acknowledgments}
We acknowledge to Prof. Y. Iwasa, Dr. T. Takenobu and S. Moriyama for
the suggestions for synthesis and heat capacity measurements. We also
thank to Prof. A. Asamitsu for the advice of the resistivity
measurements at low temperature. This work was supported by ``Research
for the Future'' of Japan Society for the Promotion of Science (JSPS),
Japan.
\end{acknowledgments}

\bibliography{eu}

\begin{thebibliography}{29}
\expandafter\ifx\csname natexlab\endcsname\relax\def\natexlab#1{#1}\fi
\expandafter\ifx\csname bibnamefont\endcsname\relax
  \def\bibnamefont#1{#1}\fi
\expandafter\ifx\csname bibfnamefont\endcsname\relax
  \def\bibfnamefont#1{#1}\fi
\expandafter\ifx\csname citenamefont\endcsname\relax
  \def\citenamefont#1{#1}\fi
\expandafter\ifx\csname url\endcsname\relax
  \def\url#1{\texttt{#1}}\fi
\expandafter\ifx\csname urlprefix\endcsname\relax\def\urlprefix{URL }\fi
\providecommand{\bibinfo}[2]{#2}
\providecommand{\eprint}[2][]{\url{#2}}

\bibitem[{\citenamefont{Hebard et~al.}(1991)\citenamefont{Hebard, Rosseinsky,
  Haddon, Murphy, Glarum, Palstra, Ramirez, and Kortan}}]{Hebard1}
\bibinfo{author}{\bibfnamefont{A.~F.} \bibnamefont{Hebard}},
  \bibinfo{author}{\bibfnamefont{M.~J.} \bibnamefont{Rosseinsky}},
  \bibinfo{author}{\bibfnamefont{R.~C.} \bibnamefont{Haddon}},
  \bibinfo{author}{\bibfnamefont{D.~W.} \bibnamefont{Murphy}},
  \bibinfo{author}{\bibfnamefont{S.~H.} \bibnamefont{Glarum}},
  \bibinfo{author}{\bibfnamefont{T.~T.~M.} \bibnamefont{Palstra}},
  \bibinfo{author}{\bibfnamefont{A.~P.} \bibnamefont{Ramirez}},
  \bibnamefont{and} \bibinfo{author}{\bibfnamefont{A.~R.}
  \bibnamefont{Kortan}}, \bibinfo{journal}{Nature (London)}
  \textbf{\bibinfo{volume}{350}}, \bibinfo{pages}{600} (\bibinfo{year}{1991}).

\bibitem[{\citenamefont{Allemand et~al.}(1991)\citenamefont{Allemand, Khemani,
  Koch, Wudl, Holczer, Donovan, Gr{\"u}ner, and Thompson}}]{Allemand1}
\bibinfo{author}{\bibfnamefont{P.~M.} \bibnamefont{Allemand}},
  \bibinfo{author}{\bibfnamefont{K.~C.} \bibnamefont{Khemani}},
  \bibinfo{author}{\bibfnamefont{A.}~\bibnamefont{Koch}},
  \bibinfo{author}{\bibfnamefont{F.}~\bibnamefont{Wudl}},
  \bibinfo{author}{\bibfnamefont{K.}~\bibnamefont{Holczer}},
  \bibinfo{author}{\bibfnamefont{S.}~\bibnamefont{Donovan}},
  \bibinfo{author}{\bibfnamefont{G.}~\bibnamefont{Gr{\"u}ner}},
  \bibnamefont{and} \bibinfo{author}{\bibfnamefont{J.~D.}
  \bibnamefont{Thompson}}, \bibinfo{journal}{Science}
  \textbf{\bibinfo{volume}{253}}, \bibinfo{pages}{301} (\bibinfo{year}{1991}).

\bibitem[{\citenamefont{Chauvet et~al.}(1994)\citenamefont{Chauvet, Oszl\`anyi,
  Forro, Stephens, Tegze, Faigel, and J\`anossy}}]{Chauvet1}
\bibinfo{author}{\bibfnamefont{O.}~\bibnamefont{Chauvet}},
  \bibinfo{author}{\bibfnamefont{G.}~\bibnamefont{Oszl\`anyi}},
  \bibinfo{author}{\bibfnamefont{L.}~\bibnamefont{Forro}},
  \bibinfo{author}{\bibfnamefont{P.~W.} \bibnamefont{Stephens}},
  \bibinfo{author}{\bibfnamefont{M.}~\bibnamefont{Tegze}},
  \bibinfo{author}{\bibfnamefont{G.}~\bibnamefont{Faigel}}, \bibnamefont{and}
  \bibinfo{author}{\bibfnamefont{A.}~\bibnamefont{J\`anossy}},
  \bibinfo{journal}{Phys. Rev. Lett.} \textbf{\bibinfo{volume}{72}},
  \bibinfo{pages}{2721} (\bibinfo{year}{1994}).

\bibitem[{\citenamefont{Ar{\v c}on et~al.}(2000)\citenamefont{Ar{\v c}on,
  Prassides, Maniero, and Brunel}}]{Arcon1}
\bibinfo{author}{\bibfnamefont{D.}~\bibnamefont{Ar{\v c}on}},
  \bibinfo{author}{\bibfnamefont{K.}~\bibnamefont{Prassides}},
  \bibinfo{author}{\bibfnamefont{A.~L.} \bibnamefont{Maniero}},
  \bibnamefont{and} \bibinfo{author}{\bibfnamefont{L.~C.}
  \bibnamefont{Brunel}}, \bibinfo{journal}{Phys. Rev. Lett.}
  \textbf{\bibinfo{volume}{84}}, \bibinfo{pages}{562} (\bibinfo{year}{2000}).

\bibitem[{\citenamefont{Takenobu et~al.}(2000)\citenamefont{Takenobu, Muro,
  Iwasa, and Mitani}}]{Takenobu1}
\bibinfo{author}{\bibfnamefont{T.}~\bibnamefont{Takenobu}},
  \bibinfo{author}{\bibfnamefont{T.}~\bibnamefont{Muro}},
  \bibinfo{author}{\bibfnamefont{Y.}~\bibnamefont{Iwasa}}, \bibnamefont{and}
  \bibinfo{author}{\bibfnamefont{T.}~\bibnamefont{Mitani}},
  \bibinfo{journal}{Phys. Rev. Lett.} \textbf{\bibinfo{volume}{85}},
  \bibinfo{pages}{381} (\bibinfo{year}{2000}).

\bibitem[{\citenamefont{{\"O}zda{\c s} et~al.}(1995)\citenamefont{{\"O}zda{\c
  s}, Kortan, Kopylov, Ramirez, Siegrist, M.Rabe, Bair, S.Schuppler, and
  Citrin}}]{Oezdas1}
\bibinfo{author}{\bibfnamefont{E.}~\bibnamefont{{\"O}zda{\c s}}},
  \bibinfo{author}{\bibfnamefont{A.~R.} \bibnamefont{Kortan}},
  \bibinfo{author}{\bibfnamefont{N.}~\bibnamefont{Kopylov}},
  \bibinfo{author}{\bibfnamefont{A.~P.} \bibnamefont{Ramirez}},
  \bibinfo{author}{\bibfnamefont{T.}~\bibnamefont{Siegrist}},
  \bibinfo{author}{\bibfnamefont{K.}~\bibnamefont{M.Rabe}},
  \bibinfo{author}{\bibfnamefont{H.~E.} \bibnamefont{Bair}},
  \bibinfo{author}{\bibnamefont{S.Schuppler}}, \bibnamefont{and}
  \bibinfo{author}{\bibfnamefont{P.~H.} \bibnamefont{Citrin}},
  \bibinfo{journal}{Nature (London)} \textbf{\bibinfo{volume}{375}},
  \bibinfo{pages}{126} (\bibinfo{year}{1995}).

\bibitem[{\citenamefont{Chen and Roth}(1995)}]{Chen1}
\bibinfo{author}{\bibfnamefont{X.~H.} \bibnamefont{Chen}} \bibnamefont{and}
  \bibinfo{author}{\bibfnamefont{G.}~\bibnamefont{Roth}},
  \bibinfo{journal}{Phys. Rev. B} \textbf{\bibinfo{volume}{52}},
  \bibinfo{pages}{15534} (\bibinfo{year}{1995}).

\bibitem[{\citenamefont{Yoshikawa et~al.}(1995)\citenamefont{Yoshikawa,
  Kuroshima, Hirosawa, Tanigaki, , and Mizuki}}]{Yoshikawa1}
\bibinfo{author}{\bibfnamefont{H.}~\bibnamefont{Yoshikawa}},
  \bibinfo{author}{\bibfnamefont{S.}~\bibnamefont{Kuroshima}},
  \bibinfo{author}{\bibfnamefont{I.}~\bibnamefont{Hirosawa}},
  \bibinfo{author}{\bibfnamefont{K.}~\bibnamefont{Tanigaki}}, ,
  \bibnamefont{and} \bibinfo{author}{\bibfnamefont{J.}~\bibnamefont{Mizuki}},
  \bibinfo{journal}{Chem. Phys. Lett.} \textbf{\bibinfo{volume}{239}},
  \bibinfo{pages}{103} (\bibinfo{year}{1995}).

\bibitem[{\citenamefont{Ksari-Habiles et~al.}(1997)\citenamefont{Ksari-Habiles,
  Claves, Chouteau, Touzain, Jeandey, Oddoou, and Stepanov}}]{Ksari1}
\bibinfo{author}{\bibfnamefont{Y.}~\bibnamefont{Ksari-Habiles}},
  \bibinfo{author}{\bibfnamefont{D.}~\bibnamefont{Claves}},
  \bibinfo{author}{\bibfnamefont{G.}~\bibnamefont{Chouteau}},
  \bibinfo{author}{\bibfnamefont{P.}~\bibnamefont{Touzain}},
  \bibinfo{author}{\bibfnamefont{C.}~\bibnamefont{Jeandey}},
  \bibinfo{author}{\bibfnamefont{J.~L.} \bibnamefont{Oddoou}},
  \bibnamefont{and} \bibinfo{author}{\bibfnamefont{A.}~\bibnamefont{Stepanov}},
  \bibinfo{journal}{J. Phys. Chem. Solids} \textbf{\bibinfo{volume}{58}},
  \bibinfo{pages}{1771} (\bibinfo{year}{1997}).

\bibitem[{\citenamefont{Claves et~al.}(1998)\citenamefont{Claves,
  Ksari-Habiles, Chouteau, and Touzain}}]{Claves1}
\bibinfo{author}{\bibfnamefont{D.}~\bibnamefont{Claves}},
  \bibinfo{author}{\bibfnamefont{Y.}~\bibnamefont{Ksari-Habiles}},
  \bibinfo{author}{\bibfnamefont{G.}~\bibnamefont{Chouteau}}, \bibnamefont{and}
  \bibinfo{author}{\bibfnamefont{P.}~\bibnamefont{Touzain}},
  \bibinfo{journal}{Solid State Commun.} \textbf{\bibinfo{volume}{106}},
  \bibinfo{pages}{431} (\bibinfo{year}{1998}).

\bibitem[{\citenamefont{Fujiwara et~al.}(2000)\citenamefont{Fujiwara, Ishii,
  Watanuki, Suematsu, Nakao, Ohwada, Fujii, Murakami, Mori, Kawada
  et~al.}}]{Fujiwara1}
\bibinfo{author}{\bibfnamefont{A.}~\bibnamefont{Fujiwara}},
  \bibinfo{author}{\bibfnamefont{K.}~\bibnamefont{Ishii}},
  \bibinfo{author}{\bibfnamefont{T.}~\bibnamefont{Watanuki}},
  \bibinfo{author}{\bibfnamefont{H.}~\bibnamefont{Suematsu}},
  \bibinfo{author}{\bibfnamefont{H.}~\bibnamefont{Nakao}},
  \bibinfo{author}{\bibfnamefont{K.}~\bibnamefont{Ohwada}},
  \bibinfo{author}{\bibfnamefont{Y.}~\bibnamefont{Fujii}},
  \bibinfo{author}{\bibfnamefont{Y.}~\bibnamefont{Murakami}},
  \bibinfo{author}{\bibfnamefont{T.}~\bibnamefont{Mori}},
  \bibinfo{author}{\bibfnamefont{H.}~\bibnamefont{Kawada}},
  \bibnamefont{et~al.}, \bibinfo{journal}{J. Appl. Cryst.}
  \textbf{\bibinfo{volume}{33}}, \bibinfo{pages}{1241} (\bibinfo{year}{2000}).

\bibitem[{\citenamefont{Zhou et~al.}(1991)\citenamefont{Zhou, Fisher, Coustel,
  Kycia, Zhu, McGhie, Romanow, Jr., III, and Cox}}]{Zhou1}
\bibinfo{author}{\bibfnamefont{O.}~\bibnamefont{Zhou}},
  \bibinfo{author}{\bibfnamefont{J.~E.} \bibnamefont{Fisher}},
  \bibinfo{author}{\bibfnamefont{N.}~\bibnamefont{Coustel}},
  \bibinfo{author}{\bibfnamefont{S.}~\bibnamefont{Kycia}},
  \bibinfo{author}{\bibfnamefont{Q.}~\bibnamefont{Zhu}},
  \bibinfo{author}{\bibfnamefont{A.~R.} \bibnamefont{McGhie}},
  \bibinfo{author}{\bibfnamefont{W.~J.} \bibnamefont{Romanow}},
  \bibinfo{author}{\bibfnamefont{J.~P.~M.} \bibnamefont{Jr.}},
  \bibinfo{author}{\bibfnamefont{A.~B.~S.} \bibnamefont{III}},
  \bibnamefont{and} \bibinfo{author}{\bibfnamefont{D.~E.} \bibnamefont{Cox}},
  \bibinfo{journal}{Nature (London)} \textbf{\bibinfo{volume}{351}},
  \bibinfo{pages}{462} (\bibinfo{year}{1991}).

\bibitem[{\citenamefont{Kortan et~al.}(1991)\citenamefont{Kortan, Kopylov,
  Glarum, Gyorgy, Ramirez, Fleming, Zhou, Thiel, Trevor, and Haddon}}]{Kortan1}
\bibinfo{author}{\bibfnamefont{A.~R.} \bibnamefont{Kortan}},
  \bibinfo{author}{\bibfnamefont{N.}~\bibnamefont{Kopylov}},
  \bibinfo{author}{\bibfnamefont{S.}~\bibnamefont{Glarum}},
  \bibinfo{author}{\bibfnamefont{E.~M.} \bibnamefont{Gyorgy}},
  \bibinfo{author}{\bibfnamefont{A.~P.} \bibnamefont{Ramirez}},
  \bibinfo{author}{\bibfnamefont{R.~M.} \bibnamefont{Fleming}},
  \bibinfo{author}{\bibfnamefont{O.}~\bibnamefont{Zhou}},
  \bibinfo{author}{\bibfnamefont{F.~A.} \bibnamefont{Thiel}},
  \bibinfo{author}{\bibfnamefont{P.~L.} \bibnamefont{Trevor}},
  \bibnamefont{and} \bibinfo{author}{\bibfnamefont{R.~C.}
  \bibnamefont{Haddon}}, \bibinfo{journal}{Nature (London)}
  \textbf{\bibinfo{volume}{360}}, \bibinfo{pages}{566} (\bibinfo{year}{1991}).

\bibitem[{\citenamefont{Chen et~al.}(1999)\citenamefont{Chen, Liu, Li, Chi, and
  Iwasa}}]{Chen2}
\bibinfo{author}{\bibfnamefont{X.~H.} \bibnamefont{Chen}},
  \bibinfo{author}{\bibfnamefont{Z.~S.} \bibnamefont{Liu}},
  \bibinfo{author}{\bibfnamefont{S.~Y.} \bibnamefont{Li}},
  \bibinfo{author}{\bibfnamefont{D.~H.} \bibnamefont{Chi}}, \bibnamefont{and}
  \bibinfo{author}{\bibfnamefont{Y.}~\bibnamefont{Iwasa}},
  \bibinfo{journal}{Phys. Rev. B} \textbf{\bibinfo{volume}{60}},
  \bibinfo{pages}{6183} (\bibinfo{year}{1999}).

\bibitem[{\citenamefont{Izumi}(1993)}]{Izumi1}
\bibinfo{author}{\bibfnamefont{F.}~\bibnamefont{Izumi}}, in
  \emph{\bibinfo{booktitle}{The Rietveld Method}}, edited by
  \bibinfo{editor}{\bibfnamefont{R.~A.} \bibnamefont{Young}}
  (\bibinfo{publisher}{Oxford University Press}, \bibinfo{address}{Oxford},
  \bibinfo{year}{1993}), chap.~\bibinfo{chapter}{13}.

\bibitem[{\citenamefont{Kim and Izumi}(1994)}]{Kim1}
\bibinfo{author}{\bibfnamefont{Y.~I.} \bibnamefont{Kim}} \bibnamefont{and}
  \bibinfo{author}{\bibfnamefont{F.}~\bibnamefont{Izumi}}, \bibinfo{journal}{J.
  Ceram. Soc. Jpn.} \textbf{\bibinfo{volume}{102}}, \bibinfo{pages}{401}
  (\bibinfo{year}{1994}).

\bibitem[{\citenamefont{Ootoshi et~al.}(2000)\citenamefont{Ootoshi, Ishii,
  Fujiwara, Watanuki, Matsuoka, and Suematsu}}]{Ootoshi1}
\bibinfo{author}{\bibfnamefont{H.}~\bibnamefont{Ootoshi}},
  \bibinfo{author}{\bibfnamefont{K.}~\bibnamefont{Ishii}},
  \bibinfo{author}{\bibfnamefont{A.}~\bibnamefont{Fujiwara}},
  \bibinfo{author}{\bibfnamefont{T.}~\bibnamefont{Watanuki}},
  \bibinfo{author}{\bibfnamefont{Y.}~\bibnamefont{Matsuoka}}, \bibnamefont{and}
  \bibinfo{author}{\bibfnamefont{H.}~\bibnamefont{Suematsu}},
  \bibinfo{journal}{Mol Cryst. and Liq. Cryst.} \textbf{\bibinfo{volume}{340}},
  \bibinfo{pages}{565} (\bibinfo{year}{2000}).

\bibitem[{eum()}]{eumag}
\bibinfo{note}{In Eu$_3$Sr$_3$C60 and Eu$_1$Sr$_5$C60, the values of
  $\mu_{eff}$ and $M_S$ in Tabel \ref{tab:magnetism} are a little larger than
  the theoretical values of Eu$^{2+}$. The most probable reason is the
  discrepancy of the stoichiometry, but the excess from the theoretical value
  is often observed in the samples of different batches, including
  Eu$_6$C$_{60}$.}

\bibitem[{\citenamefont{Inoue et~al.}(2000)\citenamefont{Inoue, Kubozono,
  Kashino, Takabayashi, Fujitaka, Hida, Inoue, Kambara, Emura, and
  Uruga}}]{Inoue1}
\bibinfo{author}{\bibfnamefont{T.}~\bibnamefont{Inoue}},
  \bibinfo{author}{\bibfnamefont{Y.}~\bibnamefont{Kubozono}},
  \bibinfo{author}{\bibfnamefont{S.}~\bibnamefont{Kashino}},
  \bibinfo{author}{\bibfnamefont{Y.}~\bibnamefont{Takabayashi}},
  \bibinfo{author}{\bibfnamefont{K.}~\bibnamefont{Fujitaka}},
  \bibinfo{author}{\bibfnamefont{M.}~\bibnamefont{Hida}},
  \bibinfo{author}{\bibfnamefont{M.}~\bibnamefont{Inoue}},
  \bibinfo{author}{\bibfnamefont{T.}~\bibnamefont{Kambara}},
  \bibinfo{author}{\bibfnamefont{S.}~\bibnamefont{Emura}}, \bibnamefont{and}
  \bibinfo{author}{\bibfnamefont{T.}~\bibnamefont{Uruga}},
  \bibinfo{journal}{Chem. Phys. Lett.} \textbf{\bibinfo{volume}{316}},
  \bibinfo{pages}{381} (\bibinfo{year}{2000}).

\bibitem[{\citenamefont{Arrott}(1957)}]{Arrott1}
\bibinfo{author}{\bibfnamefont{A.}~\bibnamefont{Arrott}},
  \bibinfo{journal}{Phys. Rev.} \textbf{\bibinfo{volume}{108}},
  \bibinfo{pages}{1394} (\bibinfo{year}{1957}).

\bibitem[{\citenamefont{Saito and Oshiyama}(1993)}]{Saito1}
\bibinfo{author}{\bibfnamefont{S.}~\bibnamefont{Saito}} \bibnamefont{and}
  \bibinfo{author}{\bibfnamefont{A.}~\bibnamefont{Oshiyama}},
  \bibinfo{journal}{Phys. Rev. Lett.} \textbf{\bibinfo{volume}{71}},
  \bibinfo{pages}{121} (\bibinfo{year}{1993}).

\bibitem[{\citenamefont{Gogia et~al.}(1998)\citenamefont{Gogia, Kordatos,
  Suematsu, Tanigaki, and Prassides}}]{Gogia1}
\bibinfo{author}{\bibfnamefont{B.}~\bibnamefont{Gogia}},
  \bibinfo{author}{\bibfnamefont{K.}~\bibnamefont{Kordatos}},
  \bibinfo{author}{\bibfnamefont{H.}~\bibnamefont{Suematsu}},
  \bibinfo{author}{\bibfnamefont{K.}~\bibnamefont{Tanigaki}}, \bibnamefont{and}
  \bibinfo{author}{\bibfnamefont{K.}~\bibnamefont{Prassides}},
  \bibinfo{journal}{Phys. Rev. B} \textbf{\bibinfo{volume}{58}},
  \bibinfo{pages}{1077} (\bibinfo{year}{1998}).

\bibitem[{\citenamefont{Kasuya}(1970)}]{Kasuya2}
\bibinfo{author}{\bibfnamefont{T.}~\bibnamefont{Kasuya}},
  \bibinfo{journal}{I.B.M. J. Res. Develop.} \textbf{\bibinfo{volume}{14}},
  \bibinfo{pages}{214} (\bibinfo{year}{1970}).

\bibitem[{\citenamefont{Samokhvalov et~al.}(1967)\citenamefont{Samokhvalov,
  Loshkareva, and Bamburov}}]{Samokhvalov1}
\bibinfo{author}{\bibfnamefont{A.~A.} \bibnamefont{Samokhvalov}},
  \bibinfo{author}{\bibfnamefont{N.~N.} \bibnamefont{Loshkareva}},
  \bibnamefont{and} \bibinfo{author}{\bibfnamefont{V.~G.}
  \bibnamefont{Bamburov}}, \bibinfo{journal}{Sov. Phys. Solid State}
  \textbf{\bibinfo{volume}{9}}, \bibinfo{pages}{555} (\bibinfo{year}{1967}).

\bibitem[{\citenamefont{de~Gennes and Friedel}(1958)}]{Gennes1}
\bibinfo{author}{\bibfnamefont{P.~G.~E.} \bibnamefont{de~Gennes}}
  \bibnamefont{and} \bibinfo{author}{\bibfnamefont{J.}~\bibnamefont{Friedel}},
  \bibinfo{journal}{J. of Phys. Chem. Solids} \textbf{\bibinfo{volume}{4}},
  \bibinfo{pages}{71} (\bibinfo{year}{1958}).

\bibitem[{\citenamefont{Fisher and Langer}(1968)}]{Fisher1}
\bibinfo{author}{\bibfnamefont{M.~E.} \bibnamefont{Fisher}} \bibnamefont{and}
  \bibinfo{author}{\bibfnamefont{J.~S.} \bibnamefont{Langer}},
  \bibinfo{journal}{Phys. Rev. Lett.} \textbf{\bibinfo{volume}{20}},
  \bibinfo{pages}{665} (\bibinfo{year}{1968}).

\bibitem[{\citenamefont{Helman and Abeles}(1976)}]{Helman1}
\bibinfo{author}{\bibfnamefont{J.~S.} \bibnamefont{Helman}} \bibnamefont{and}
  \bibinfo{author}{\bibfnamefont{B.}~\bibnamefont{Abeles}},
  \bibinfo{journal}{Phys. Rev. Lett.} \textbf{\bibinfo{volume}{37}},
  \bibinfo{pages}{1429} (\bibinfo{year}{1976}).

\bibitem[{\citenamefont{Sheng et~al.}(1972)\citenamefont{Sheng, Abeles, and
  Arie}}]{Sheng1}
\bibinfo{author}{\bibfnamefont{P.}~\bibnamefont{Sheng}},
  \bibinfo{author}{\bibfnamefont{B.}~\bibnamefont{Abeles}}, \bibnamefont{and}
  \bibinfo{author}{\bibfnamefont{Y.}~\bibnamefont{Arie}},
  \bibinfo{journal}{Phys. Rev. Lett.} \textbf{\bibinfo{volume}{31}},
  \bibinfo{pages}{44} (\bibinfo{year}{1972}).

\bibitem[{\citenamefont{Kasuya and Yanase}(1968)}]{Kasuya1}
\bibinfo{author}{\bibfnamefont{T.}~\bibnamefont{Kasuya}} \bibnamefont{and}
  \bibinfo{author}{\bibfnamefont{A.}~\bibnamefont{Yanase}},
  \bibinfo{journal}{Rev. Mod. Phys.} \textbf{\bibinfo{volume}{40}},
  \bibinfo{pages}{684} (\bibinfo{year}{1968}).

\end{thebibliography}

\newpage

\begin{table}
\caption{\label{tab:structure} Structural parameter obtained from the
Rietveld refinement of Eu$_{6-x}$Sr$_x$C$_{60}$. }
\begin{ruledtabular}
\begin{tabular}{ccccccc}
\multicolumn{7}{c}{Eu$_6$C$_{60}$ $a_0$ = 10.940 $\pm$ 0.001 \AA\
$R_{wp}$ = 3.81 \%}\\
&Site&Occupancy&$x$&$y$&$z$&$B$(\AA$^{2}$)\\
\colrule
C1&$24g$&1&0.0672(5)&0&0.3200&0.5(2)\\
C2&$48h$&1&0.1325&0.1056&0.2797&0.5\\
C3&$48h$&1&0.0653(3)&0.2144&0.2381&0.5\\
Eu$^{2+}$&$12e$&1&0&0.5&0.2768(2)&2.29(4)\\
\end{tabular}
\begin{tabular}{ccccccc}
\multicolumn{7}{c}{Eu$_3$Sr$_3$C$_{60}$ $a_0$ = 10.958 $\pm$ 0.002 \AA\
$R_{wp}$ = 6.20 \%}\\
&Site&Occupancy&$x$&$y$&$z$&$B$(\AA$^{2}$)\\
\colrule
C1&$24g$&1&0.0686(6)&0&0.3186&1.1(3)\\
C2&$48h$&1&0.1328&0.1038&0.2790&1.1\\
C3&$48h$&1&0.0641(4)&0.2148&0.2366&1.1\\
Eu$^{2+}$&$12e$&0.51(2)&0&0.5&0.2792(3)&2.99(7)\\
Sr$^{2+}$&$12e$&0.49&0&0.5&0.2792&2.99\\
\end{tabular}
\begin{tabular}{ccccccc}
\multicolumn{7}{c}{Eu$_1$Sr$_5$C$_{60}$ $a_0$ = 10.971 $\pm$ 0.002 \AA\
$R_{wp}$ = 6.37 \%}\\
&Site&Occupancy&$x$&$y$&$z$&$B$(\AA$^{2}$)\\
\colrule
C1&$24g$&1&0.0660(5)&0&0.3207&2.1(3)\\
C2&$48h$&1&0.1321&0.1069&0.2798&2.1\\
C3&$48h$&1&0.0661(3)&0.2138&0.2390&2.1\\
Eu$^{2+}$&$12e$&0.18(2)&0&0.5&0.2803(3)&2.73(6)\\
Sr$^{2+}$&$12e$&0.82&0&0.5&0.2803(3)&2.73\\
\end{tabular}
\begin{tabular}{ccccccc}
\multicolumn{7}{c}{Sr$_6$C$_{60}$ $a_0$ = 10.986 $\pm$ 0.002 \AA}\\
\end{tabular}
\end{ruledtabular}
\end{table}

\begin{table}
\caption{\label{tab:magnetism} Summary of the magnetic properties of
Eu$_{6-x}$Sr$_x$C$_{60}$. $\mu_{eff}$, $\Theta$, $M_S$, and $T_C$
denote effective Bohr magneton ($g_J\sqrt{J(J+1)}\mu_B$),
Weiss temperature, saturation moment, and ferromagnetic transition
temperature, respectively.}
\begin{ruledtabular}
\begin{tabular}{ccccc}
&$\mu_{eff}$/Eu ($\mu_B$)&$\Theta$ (K)&$M_S$/Eu ($\mu_B$)&$T_C$ (K)\\
\colrule
Eu$^{2+}$&7.94&&7&\\
Eu$_6$C$_{60}$&7.77&10.6&6.97&11.6\footnotemark[1]\\
&&&&(13.7\footnotemark[2])\\
Eu$_3$Sr$_3$C$_{60}$&8.13&12.6&7.73&12.8\footnotemark[2]\\
Eu$_1$Sr$_5$C$_{60}$&8.26&8.0&7.68&10.4\footnotemark[2]\\
\end{tabular}
\end{ruledtabular}
\footnotetext[1]{From the peak position of heat capacity measurement.}
\footnotetext[2]{From the Arrott plot.}
\end{table}

\end{document}